\documentstyle[12pt]{article}

\begin{document}
\title{\bf {Generalized Second Law of Thermodynamics in Quintom Dominated
Universe}}
\author{M.R. Setare  \footnote{E-mail: rezakord@ipm.ir}
  \\{Department of Science, University of Kurdistan, Sanandaj, Iran}}

 \maketitle
\begin {abstract}
In this paper we will investigate the validity of the Generalized
Second Law of thermodynamics for the Quintom model of dark
energy. Reviewing briefly the quintom scenario of dark energy, we
will study the conditions of validity of the generalized second
law of thermodynamics in three cases:  quintessence dominated,
phantom dominated and transition from quintessence to phantom
will be discussed.
\end {abstract}
\newpage
\section{Introduction}
One of the most important problems of cosmology, is the problem of
so-called dark energy (DE). The type Ia supernova observations
suggests that the universe is dominated by dark energy with negative
pressure which provides the dynamical mechanism of the accelerating
expansion of the universe \cite{{per},{gar},{ries}}. The strength of
this acceleration is presently matter of debate, mainly because it
depends on the theoretical model implied when interpreting the data.
Most of these models are based on dynamics of a scalar or
multi-scalar fields (e.g quintessence \cite{{rat},{zlat}} and
quintom model of dark energy, respectively). Primary scalar field
candidate for dark energy was quintessence scenario, a fluid with
the parameter of the equation of state lying in the range, $-1< w<
{-1 \over 3}$. While the most model independent analysis suggest
that the acceleration of the universe to be below the de Sitter
value \cite{Daly}, it is certainly true that the body of
observational data allows for a wide parameter space compatible with
an acceleration larger than the de Sitter's \cite{{cal},{hans}}. If
eventually this proves to be the case, the fluid driving the
expansion would violate not only the strong energy condition $\rho +
3P>0$, but the dominate energy condition $\rho + P>0$, as well.
Fluids of such characteristic dubbed phantom fluid \cite{cal}. In
spite of the fact that the field theory of phantom fields encounter
the problem of stability which one could try to bypass by assuming
them to be effective fields \cite{{car},{gib}}, it is nevertheless
interesting to study their cosmological implication. Recently there
are many relevant studies on phantom energy \cite{meng}. The
analysis of the properties of dark energy from recent observations
mildly favor models with $w$ crossing -1 in the near past. But,
neither quintessence nor phantom can fulfill this transition. In the
quintessence model, the equation of state $w=p/\rho$ is always in
the range $-1\leq w\leq 1$ for $V(\phi)>0$. Meanwhile for the
phantom which has the opposite sign of the kinetic term compared
with the quintessence in the Lagrangian, one always has $w\leq -1$.
Neither the quintessence nor the phantom alone can fulfill the
transition from $w>-1$ to $w<-1$ and vice versa. Although for
k-essence\cite{k-essence} one can have both $w\ge -1$ and $w<-1$, it
has been lately considered by Ref\cite{Vikman1, Vikman2} that it is
very difficult for k-essence to get $w$ across $-1$ during evolving.
But one can show \cite{{quintom},{quint2}} that considering the
combination of quintessence and phantom in a joint model, the
transition can be fulfilled. This model, dubbed quintom, can produce
a better fit to the data than more familiar models with $w\geq-1$.
In the other term the quintom model of dark energy represents a
transition of dark energy equation of state from $w>-1$ to $w<-1$,
or vice versa, namely from $w<-1$ to $w>-1$ is also one realization
of quintom, as can be seen clearly in \cite{ref}.  We must mention
that there are another possibilities in model building regarding
quintom, see \cite{sing} for a dark energy model which includes
higher derivative operators in the Lagrangian with a single scalar
field which gives rise to an equation of state larger than $-1$ in
the past and less than $-1$ at the present time. One another is the
scalar field model with non-minimal coupling to the gravity
\cite{far}. Another model is a single scalar field minimally coupled
to the gravity but with a non-minimal kinetic term, in this model a
dimensionless function of temperature $f(T)$ is in front of the
kinetic term. During the evolution of the universe when $f(T)$
change sign from positive to negative, the possibility of
quintessence-phantom transition appears \cite{quintom}.

\par
 In 1973, Bekenstein \cite{bek} assumed that
there is a relation between the event of horizon and the
thermodynamics of a black hole, so that the event of horizon of
the black hole is a measure of the entropy of it. This idea has
been generalized to horizons of cosmological models, so that each
horizon corresponds to an entropy. Thus the second law of
thermodynamics was modified in the way that in generalized form,
the sum of all time derivative of entropies related to horizons
plus time derivative of normal entropy must be positive i.e. the
sum of entropies must be increasing function of time. In
\cite{davies2}, the validity of Generalized Second Law (GSL) for
the cosmological models which departs slightly from de Sitter
space is investigated. However, it is only natural to associate
an entropy to the horizon area as it measures our lack of
knowledge about what is going on beyond it. In this paper we show
that the sum of normal entropy and the horizon entropy in phantom
dominated universe is non-decreasing function of time. Also, the
transition from quintessence to Phantom dominated universe is
considered and the conditions of the validity of GSL in transition
is studied. Also for quintom model of dark energy \cite{quintom},
we study the GSL in quintom dominated universe and conclude the
same results when we consider two scalar fields with no coupling
potential term. In our calculations we use $c=8\pi G_N=1$. \qquad
\section{The quintom model of dark energy }
The quintom model of dark energy \cite{quintom} is of new models
proposed to explain the new astrophysical data, due to transition
from $w>-1$ to $w<-1$, i.e. transition from quintessence dominated
universe to phantom dominated universe. Here we consider the
spatially flat Friedman-Robertson-Walker universe, where has
following space-time metric
\begin{equation}\label{met}
ds^{2}=-dt^{2}+a(t)^{2}(dr^{2}+r^{2}d\Omega^{2})
\end{equation}
Containing the normal scalar field $\sigma$ and negative kinetic
scalar field $\phi$, the action which describes the quintom model
is expressed as the following form
\begin{equation}\label{1}
S=\int d^4x\sqrt{-g}\left(\frac{R}{2}
 -\frac{1}{2}g^{\mu \nu}\partial _\mu \phi \partial _\nu \phi
 +\frac{1}{2}g^{\mu \nu}\partial _\mu \sigma \partial _\nu \sigma
 +V(\phi ,\sigma)\right),
\end{equation}
where we have not considered the lagrangian density of matter
field. In the spatially flat Friedman-Robertson-Walker (FRW)
universe, the effective energy density, $\rho$, and the effective
pressure, P, of the scalar fields can be described by
\begin{eqnarray}\label{2}
\rho=-\frac{1}{2}\dot{\phi}^2+\frac{1}{2}\dot{\sigma}^2+V(\phi,\sigma), \\
P=-\frac{1}{2}\dot{\phi}^2+\frac{1}{2}\dot{\sigma}^2-V(\phi,\sigma).
\end{eqnarray}
So, the equation of state can be written as
\begin{equation}\label{3}
\label{ES} w=\frac{-\dot{\phi}^2+\dot{\sigma}^2-2V(\phi,\sigma)}
{-\dot{\phi}^2+\dot{\sigma}^2+2V(\phi,\sigma)}.
\end{equation}
From the equation of state , it is seen that for
$\dot\sigma>\dot\phi$, $w\geq -1$ and for $\dot\sigma<\dot\phi$,
we will have, $w<-1$. Alike \cite{quint2}, we consider a potential
with no direct coupling between two scalar fields
\begin{equation}\label{4}
V(\phi,\sigma)=V_{\phi}(\phi)+V_{\sigma}(\sigma)=V_{\phi 0}\,
 e^{-\lambda_\phi \phi}+V_{\sigma 0}\,
 e^{-\lambda_\sigma \sigma},
\end{equation}
Where the $\lambda_\phi$ and $\lambda_\sigma$, are two
dimensionless positive numbers characterizing the slope of the
potential for $\phi$ and $\sigma$ respectively. So, the evolution
equation for two scalar fields in FRW model will have the
following form
\begin{eqnarray}\label{5}
\ddot{\phi}+3H\dot{\phi}-\frac{dV_{\phi}(\phi)}{d\phi} &=& 0, \\
\ddot{\sigma}+3H\dot{\sigma}+\frac{dV_{\sigma}(\sigma)}{d\sigma}
 &=& 0,
\end{eqnarray}
where, H is the Hubble parameter. \qquad
\section{Generalized second law and quintom model of dark energy}
To study the GSL through the universe which is dominated by
quintom scenario, we deduce the expression for normal entropy
using the first law of thermodynamics.
\begin{equation}\label{6}
TdS=dE + PdV =(P+\rho)dV+Vd\rho
\end{equation}
From the equations (\ref{2}),(4) we have
\begin{equation}\label{7}
P+\rho = -{\dot\phi}^2 + {\dot\sigma}^2
\end{equation}
and the Friedman constraint equation will be
\begin{equation}\label{8}
H^2= {1\over 3}({-{\dot\phi}^2\over 2}+ V_\phi +
{{\dot\sigma}^2\over 2}+ V_\sigma ).
\end{equation}
So, using relations (\ref{5}) and (8), it is seen that
\begin{equation}\label{9}
\dot H={1\over 2}({\dot\phi}^2-{\dot\sigma}^2)={-1\over 2}(P+\rho)
\end{equation}
Thus, if ${\dot\phi}^2<{\dot\sigma}^2$ then $\dot H<0$ , i.e. for
the quintessence dominated universe and if
${\dot\phi}^2>{\dot\sigma}^2$ then $\dot H>0$, for the phantom
dominated universe. Rewriting the first law of thermodynamics
with respect to relations above and using $V={4\over3}\pi
{R_h}^3$, in which the $R_h$ is the event of horizon, one can
obtain
\begin{equation}\label{10}
TdS= -2\dot H dV+ Vd\rho= -8\pi{R_h}^2\dot H dR_h+ 8\pi{R_h}^3HdH
\end{equation}
where $T$ is the temperature of the quintom fluid. Therefore, the
time derivative of normal entropy will have the following form
\begin{equation}\label{11}
\dot S={8\pi\dot H {R_h}^2 \over T}(HR_h-\dot R_h)
\end{equation}
As we know, the quintom is the combination of normal scalar
filed, i.e, quintessence and phantom scalar field. From the
definition of event of horizon
\begin{equation}\label{12}
R_h=a(t)\int_t^{t_s}{dt'\over{a(t')}},\qquad
\int_t^{t_s}{dt'\over{a(t')}}<\infty.
\end{equation}
where for different space times $t_s$ has different values, e.g
for de Sitter space time $t_s = \infty$, $R_h$ satisfies the
following equation which is true for both scalar fields
individually
\begin{equation}\label{13}
\dot{R_h}=HR_h-1
\end{equation}
where $\dot R_h \leq0$ for phantom dominated universe
\cite{mohseni} and $\dot R_h \geq 0$ for quintessence dominated
universe \cite{davies2}. As the final form, we write the time
derivative of normal entropy of the quintom fluid using
relation(\ref{10})
\begin{equation}\label{14}
\dot{S}=\frac{8\pi\dot{H}R^{2}_{h}}{T}
\end{equation}
As it is seen from relation (14), it is shown that, the sign of
$\dot S$ depends on the sign of $\dot H$, hence for quintessence
dominated universe $\dot S<0$ and for phantom dominated universe
$\dot S>0$.
\par
The entropy of a black hole is proportional to the area of its
event horizon is well understood, it has deep physical meaning.
The status of an entropy associated to a cosmological event
horizon is not well established. In some cases like the case a de
Sitter horizon this seems plausible, with some caveats, but in
general this is a topic of current research; see \cite{pa}.
 If, the horizon entropy is taken to be $S_h=\pi
R_h^2$, the generalized second law stated that
\begin{equation}\label{15}
\dot S+ \dot S_h \geq0
\end{equation}
Thus, we will have
\begin{equation}\label{16}
\dot S+ \dot S_h = {8\pi\dot H {R}^2_{h} \over T} + 2\pi R_h \dot
R_h\geq0
\end{equation}
To investigate the validity of equation (19), we will consider
three different cases, the first case we dominate the phantom
fluid, the second, the quintessence will be dominated, and the third, the transition from quintessence to phantom.\\

a) Phantom dominated:\newline
 In this case $\dot R_h\leq0$ and $\dot H
>0$, then $\dot S_{h}<0$. If the phantom fluid temperature $T>0$ the condition for validity of GSL is as
\begin{equation}
\dot H\geq |{ T \dot R_h \over 2R_h}|
\end{equation}
If the temperature is assumed to be proportional to the de Sitter
temperature \cite{davies2}
\begin{equation}
T=\frac{bH}{2\pi} \label{detem}
\end{equation}
where $b$ is a parameter, the GSL hold when:
\begin{equation}\label{con}
b\leq\frac{4\pi\dot{H}R_h}{H|\dot R_h|}
\end{equation}
in de Sitter spacetime case $R_h=\frac{1}{H}$, then $b\leq 1$. In
phantom model case which is small perturbed around de Sitter
space, one can expect $T\leq\frac{H}{2\pi}$, which is the
condition that the phantom fluid be cooler than the horizon
temperature.
\par
b) Quintessence dominated:\newline\\
In this case $\dot R_h\geq0$, and $\dot H <0$ so the sum of normal
entropy and horizon entropy could be positive, if $T>0$ then the
condition for validity of GSL is
\begin{equation}\label{qico}
|\dot H|\leq { T \dot R_h \over 2R_h}
\end{equation}
using eq.(\ref{detem})this condition has following form
\begin{equation}\label{bcon}
b\geq \frac{4\pi|\dot{H}|R_h}{H \dot R_h}
\end{equation}
\par
 c) Phase transition from quintessence to phantom:\newline\\
As $\dot R_h\geq0 $ in quintessence model and $\dot R_h\leq0 $ in
phantom model, and assuming  that $R_h$ variates continually one can
expect that in transition from quintessence to phantom $\dot R_h=0
$. So the horizon entropy in transition time will be zero, also in
transition time $\dot{H}=0$, using eq.(\ref{14}), we obtain
$\dot{S}=0$. Therefore in the transition time the total entropy is
differentiable and continuous.
\section{Conclusion}
In order to solve cosmological problems and because the lack of our
knowledge, for instance to determine what could be the best
candidate for DE to explain the accelerated expansion of universe,
the cosmologists try to approach to best results as precise as they
can by considering all the possibilities they have. Investigating
the principles of thermodynamics and specially the second law- as
global accepted principle in the universe - in different models of
DE, as one of these possibilities, has been widely studied in the
literature, since this investigation can constrain some of
parameters in studied models, say, P. C. Davies \cite{davies2}
studied the change in event horizon area in cosmological models that
depart slightly from de Sitter space and showed that for this models
the GSL is respected for the normal scalar field, provided the fluid
to be viscous.\\
In the present paper we have considered total entropy as the entropy
of a cosmological event horizon plus the entropy of a normal scalar
field $\sigma$ and ghost scalar field $\phi$. In the quintom model
of dark energy $\dot{H}$ is given by eq.(12), for the phantom
dominated case $\dot{H}>0$, in this case $\dot{R}_{h}\leq 0 $, then
the horizon entropy is constant or decreasses with time, i.e
$\dot{S}_{h}\leq 0$, therefore the phantom entropy must increases
with expansion so long as $T>0$. In fact the phantom fluids possess
negative entropy and equals to minus the entropy of black hole of
radius $R_h$. In contrast with the previous case in the quintessence
dominated case, $\dot H <0$ and  $\dot R_h\geq0$, then
$\dot{S}_{h}\geq 0$. By considering the influence of the transition
from the quintessence to phantom dominated universe on the GSL , one
can obtain that the time derivative of the future event horizon and
the entropy must be zero at the transition time. In the summary, we
have examined the quintessence and phantom dominated universe, and
we have shown that by satisfying the conditions (20), (23) the total
entropy is non-decreasing function of time. Otherwise the  second
law of thermodynamics break down.  Note that in \cite{abdalla} these
calculations have been done for the case of interacting holographic
dark energy with dark matter, the authors have shown, in contrast to
the case of the apparent horizon,  both the first and second law of
thermodynamics break down if one consider the universe to be
enveloped by the event horizon with the usual definitions of entropy
and temperature.

\section{Acknowledgment}
The author would like to thank the referee because of his/her useful
comments, which assisted to prepare better frame for this study.

 \end{document}